\documentclass[10pt,twocolumn,letterpaper]{article}

\usepackage{cvpr}
\usepackage{times}
\usepackage{epsfig}
\usepackage{graphicx}
\usepackage{amsmath}
\usepackage{amssymb}
\usepackage{ mathrsfs }
\usepackage{multirow}

\usepackage{pifont}
\usepackage{float}
\usepackage{placeins}
\newcommand{\xmark}{\ding{55}}%

\usepackage[pagebackref=true,breaklinks=true,letterpaper=true,colorlinks,bookmarks=false]{hyperref}

\def\figvspace{{\vspace{-3mm}}}

\newcommand{\Paragraph}[1]{\vspace{1.25mm} \noindent \textbf{#1} \hspace{0mm}}
\newcommand{\Section}[1]{\vspace{-1mm} \section{#1} \vspace{-1mm}}
\newcommand{\SubSection}[1]{\vspace{-1mm} \subsection{#1} \vspace{-1mm}}
\newcommand{\SubSubSection}[1]{\vspace{-3mm} \subsubsection{#1} \vspace{-1mm}}
\newcommand{\xdownarrow}[1]{%
  {\left\downarrow\vbox to #1{}\right.\kern-\nulldelimiterspace}
}

\usepackage[pagebackref=true,breaklinks=true,letterpaper=true,colorlinks,bookmarks=false]{hyperref}

\cvprfinalcopy 

\def\cvprPaperID{12} 

\ifcvprfinal\fi
\begin{document}

\title{Color-wise Attention Network for Low-light Image Enhancement}

\author{Yousef Atoum\\
Yarmouk University\\
\and
Mao Ye\\
BOSCH Research North America\\
\and
Liu Ren\\
BOSCH Research North America\\
\and
Ying Tai\\
Tencent YouTu\\
\and
Xiaoming Liu\\
Michigan State University\\
}

\maketitle

\begin{abstract}
Absence of nearby light sources while capturing an image will degrade the visibility and quality of the captured image, making computer vision tasks difficult.
In this paper, a color-wise attention network (CWAN) is proposed for low-light image enhancement based on convolutional neural networks.
Motivated by the human visual system when looking at dark images, CWAN learns an end-to-end mapping between low-light and enhanced images while searching for any useful color cues in the low-light image to aid in the color enhancement process.
Once these regions are identified, CWAN attention will be mainly focused to synthesize these local regions, as well as the global image. 
Both quantitative and qualitative experiments on challenging datasets demonstrate the advantages of our method in comparison with state-of-the-art methods.
\end{abstract}

\Section{Introduction}

\begin{figure}[t]
\begin{center}
   \includegraphics[width=1\linewidth]{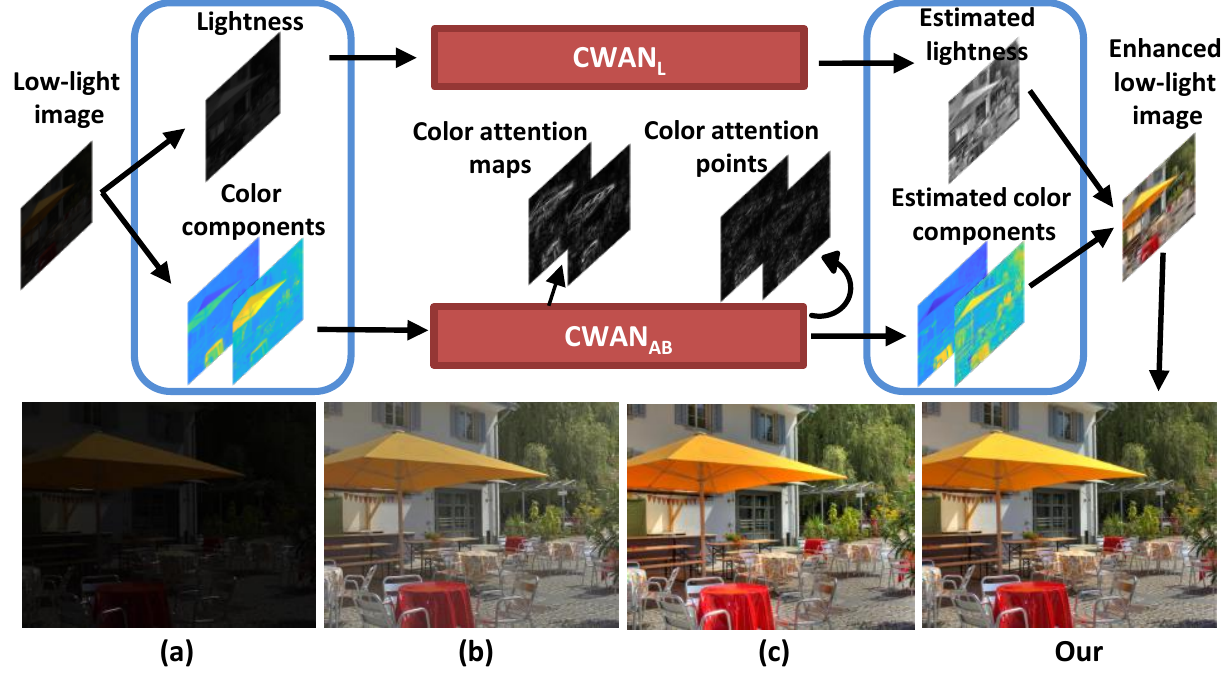}
\end{center}
\vspace{-0.5 cm}
   \caption{Our proposed CWAN method is illustrated in the top figure, where CWAN$_L$ enhances the lightness component and CWAN$_{AB}$ enhances the color components using an attention mechanism. We compare our result with (a) the low-light input, (b) LightenNet~\cite{li2018lightennet}, and (c) ground truth.}
   \vspace{-0.2 cm}
\label{fig:intro}
\end{figure}

High-quality images carry rich information of the captured scene, enabling high-level computer vision tasks, such as object detection, object recognition, and scene understanding.
However, challenges are often introduced in real-world environments making RGB-based perception difficult for both computers and humans.
Among these challenges is the low-light condition in images captured in dark environments, such as night time or dark room images. 
They suffer from degraded brightness and contrast, added noise artifacts, and have a very narrow range of colors, such that understanding the original colors in the scene is a tedious task. 
E.g., in the low-light image of Fig.~\ref{fig:intro}, many details such as the color of the building and trees are degraded and lost with the dark background.

Various methods have been proposed for the low-light image enhancement (LLIE), where researchers normally adopt a decomposition strategy for the sole purpose of simplifying the problem.
The decomposition strategy can be categorized into algorithm decomposition and image decomposition.
Algorithm decomposition includes breaking the LLIE process into separate stages.
For example, applying a denoising method prior to a lighting enhancement model~\cite{tao2017low}, or vice-versa~\cite{guo2017lime,shen2017msr}.
Many Retinex-based methods attempt to estimate the illumination component first, and then the reflectance~\cite{park2018dual}.  
In~\cite{cai2018learning}, the enhancement is decomposed into three stages: luminance enhancement, detail enhancement, and a final enhancement stage.
On the other hand, image decomposition includes breaking the image into multiple images where each holds a unique feature. 
For example, in~\cite{zhang2012enhancement,cai2016dehazenet} the original low-light image along with the decomposed inverted image are used in a dehazing algorithm.
In~\cite{ying2017bio}, they attempt to synthesize multi-exposure images to be used in a fusion algorithm.

In this paper, we propose to decompose the image into lightness and color components using the CIE LAB space, where each component is enhanced independently as seen in Fig.~\ref{fig:intro}.
Therefore, we use both algorithm and image decomposition strategies.
Our main motivation is to simplify the LLIE problem, by breaking it into two smaller problems.
One is responsible for estimating the optimal lighting condition along with denoising  given the dark image, while the other is required to revive the color information into its original state.
Another motivation is the need to pay more attention to the color components.
E.g., the estimated images from prior LLIE methods usually have degraded colors as in Fig.~\ref{fig:intro} (b).

The attention mechanism has been studied in a wide variety of computer vision problems, including object detection~\cite{yoo2015attentionnet}, tracking~\cite{kahou2015ratm}, segmentation~\cite{chen2016attention}, and action recognition~\cite{sharma2015action}.
These methods mimic the human visual system in making substantial use of contextual information in understanding RGB images.
This involves discarding unwanted regions in the image, while focusing on more important parts containing rich features of our vision task.
In this paper, we propose a novel color-wise attention CNN model, driven by key color features embedded in the low-light image.
We hypothesize that these key colors can provide useful cues for image enhancement.
These cues will be used as prior information in guiding and spanning the network's attention to faithfully recover the color of the original image.

After decomposing a low-light image into lightness and color components, each component is enhanced independently via two CNN models named CWAN$_L$ and CWAN$_{AB}$ as in Fig.~\ref{fig:intro}.
Our CWAN$_{AB}$ computes color frequencies~\cite{kashiwagi2007introduction} in the dark image, and  selects the colors with desired frequencies as the target.
We experimentally show that, by ignoring high and low frequencies, we select colors belonging to regions of interest in an image, referred to as foreground colors in the scene.
As humans, these foreground colors are what catches our eyes when first looking at dark images compared to background colors which has high frequency count.
We learn CWAN$_{AB}$ to focus its attention in enhancing points belonging to these foreground colors.
Note that the attention mechanism does not apply to CWAN$_L$, which performs lightness enhancement and denoising independently using a memory network structure~\cite{tai2017memnet}.
We demonstrate state-of-the-art (SOTA) performance on real~\cite{Chen2018SID,EKDMU17HDR} and synthetic~\cite{pascal-voc-2012} low-light datasets.

In summary, our main contributions are the following: 

$\diamond$  Propose a novel color-wise attention network (CWAN) for LLIE.
    CWAN enhances the lightness image separately from enhancing the color component.
    By doing so we simplify the LLIE problem and achieve the state of the art.
    
$\diamond$  Propose a supervised attention mechanism utilizing color frequency maps in training CWAN$_{AB}$. 
    With the color frequencies in an image, we select key local color points in the dark image which we desire CWAN to emphasize the enhancement on.
    Learning these selected colors is by nature a good starting point to spark the network's attention. 
    
$\diamond$  Achieve SOTA performance on several databases including both real and synthetic low-light images.

\Section{Related Work}

\Paragraph{Generic Low-light enhancement methods}
Low-light image enhancement has been addressed over the past decades with various handcrafted techniques.
A classical approach is to apply histogram equalization (HE), gamma correction, and their variations~\cite{pizer1987adaptive,ibrahim2007brightness}. 
Other researchers attempt more complex and global processing pipelines, e.g., estimating an illumination map via a Retinex-based method.
LIME~\cite{guo2017lime} and JED~\cite{ren2018joint} both propose Retinex-based approaches for simultaneous LLIE and noise removal.
Li et al.~\cite{li2018structure} additionally estimate a noise map in the Retinex pipeline.
AMSR~\cite{lee2013adaptive} proposes an adaptive multi-scale Retinex such that it assigns a weight to each single-scale Retinex output based on the image content. 

\begin{figure}[t!]
   \centering
\begin{tabular}{@{}c@{}c@{}c@{}c@{}}
\includegraphics[width=0.25\linewidth]{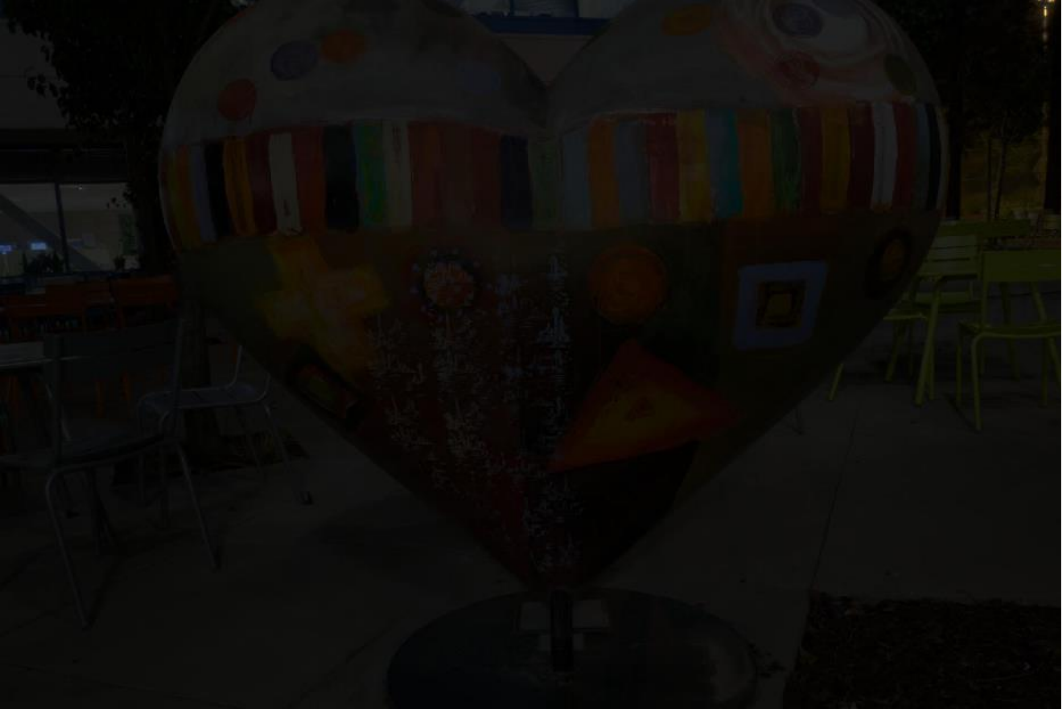}&
\includegraphics[width=0.25\linewidth]{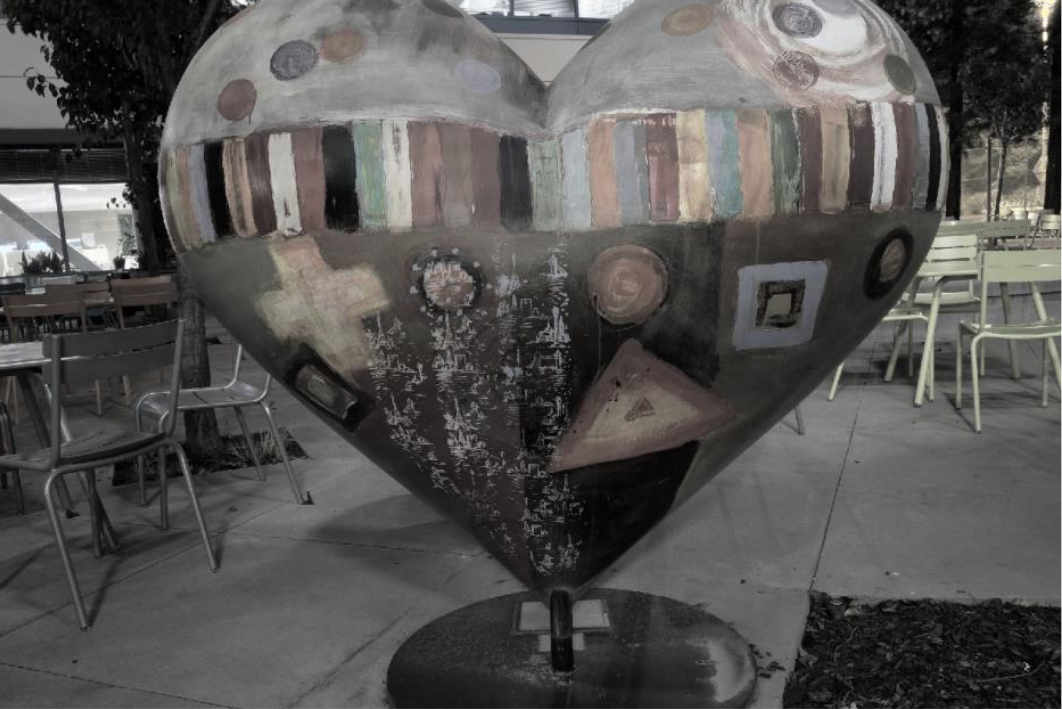}&
\includegraphics[width=0.25\linewidth]{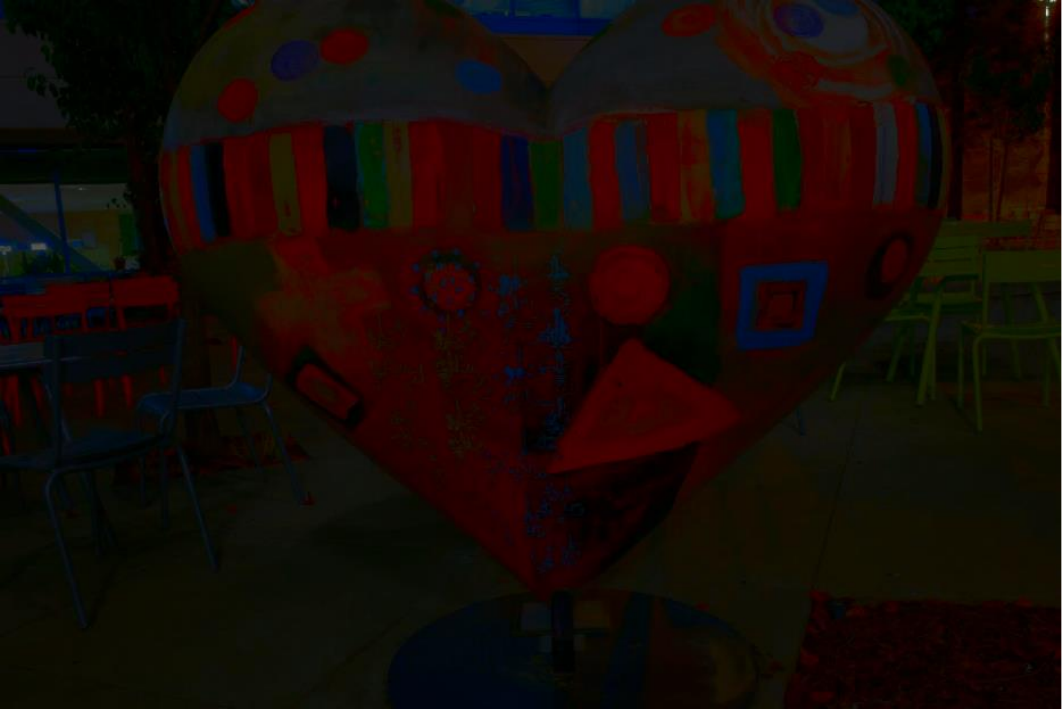}&
\includegraphics[width=0.25\linewidth]{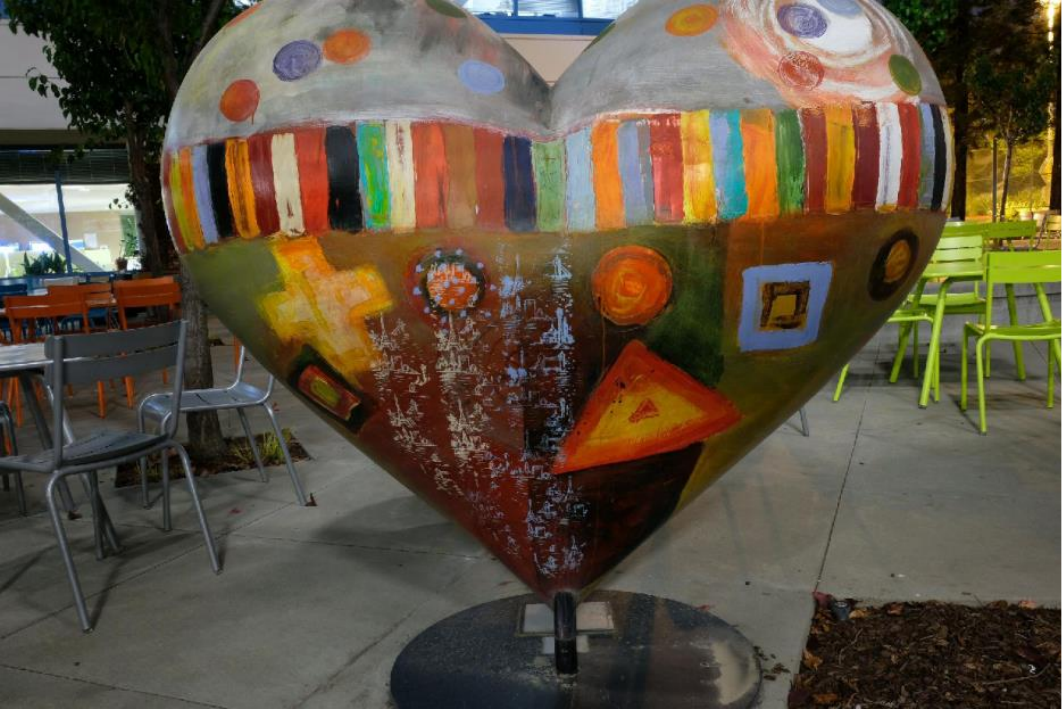}\vspace{-1mm}\\
\small{(a)}&\small{(b)}&\small{(c)}&\small{(d)}
\end{tabular}
    \caption{Decomposing an image into lightness and color components. (a) Input low-light image. (b) Lightness enhancement {\it only} using CWAN$_L$. (c) Color enhancement {\it only} using CWAN$_{AB}$. (d) The enhanced image from the proposed CWAN model.}
    \label{fig:Motivation}
    \vspace{-0.2 cm}
\end{figure}

Researchers adopt techniques from other low-level image enhancing disciplines.
For example, Ying et al.~\cite{Ying2017Neww} use the response characteristics of cameras for LLIE.
BIMEF~\cite{ying2017bio} and LECARM~\cite{ren2018lecarm} both use the camera response model to synthesize multi-exposure images for fusion.
By observing that the inverted low-light images intuitively look like haze images, the methods of~\cite{Dong2011Fast,zhang2012enhancement} apply image dehazing on the inverted image to enhance the image. 
All generic methods are based on handcrafted features and certain statistical models with many hyper-parameters.
Thus, it is difficult for these methods to work in diverse real-world scenes.

\begin{figure*}[t]
\begin{center}
   \includegraphics[width=0.85\linewidth]{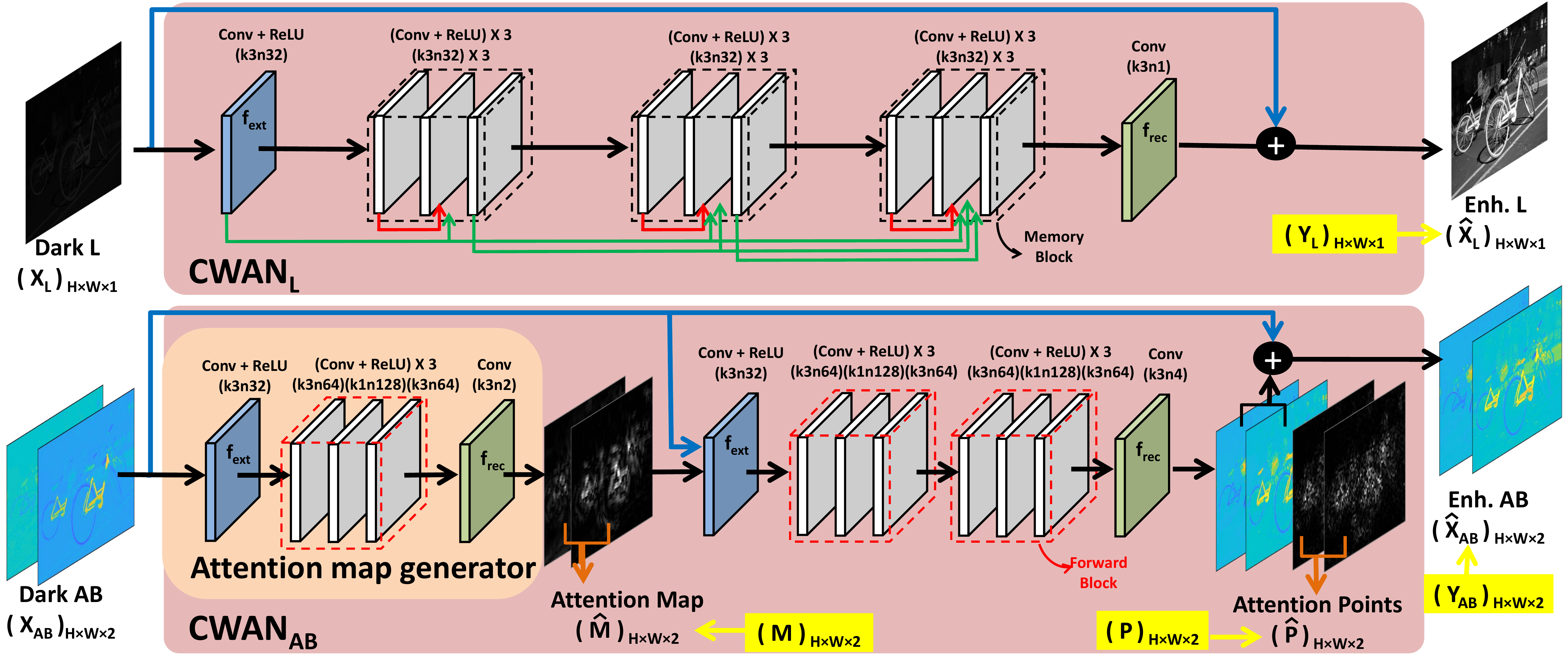}
\end{center}
\vspace{-0.4 cm}
   \caption{Network structure of the proposed CWAN method. Here, k3n64 indicates a kernel size of $3 \times 3$, and a feature map number of $64$. The stride is always equal to one in all layers. \textcolor{red}{Red}, \textcolor{green}{green} and \textcolor{blue}{blue} arrows mean short, long and global skip connection representing local short-term memory, long-term memory, and residual learning respectively. The \textcolor{yellow}{yellow} arrow represents supervision in training CWAN.}
   \vspace{-0.2 cm}
\label{fig:Arch}
\end{figure*}

\Paragraph{Low-light enhancement via CNNs}
Deep neural network has witnessed success in low-level vision problems, e.g., super-resolution~\cite{chen2018fsrnet,Tai2017disentangled}, denoising~\cite{chen2018image}, colorization~\cite{zhang2016colorful,Yoo_2019_CVPR}, dehazing~\cite{cai2016dehazenet,Qu_2019_CVPR}, multi-exposure fusion~\cite{prabhakar2017deepfuse}, and tone mapping~\cite{endo2017deep}.
CNNs have also been developed for low-light enhancement.
LLNet~\cite{lore2017llnet} proposes a stacked sparse auto-encoder, for joint LLIE and denoising, which uses gray-scale images only without considering colors.
Tao et al.~\cite{tao2017low} also propose a joint enhancement and denoising framework based on atmosphere scattering models. 
Cai et al.~\cite{cai2018learning} utilize multi-exposure fusion (MEF) datasets to learn a joint luminance and detail enhancement CNNs.
Similarly, MBLLEN~\cite{Lv2018MBLLEN} uses multiple subnets and produces the enhanced image via multi-branch fusion.
Wang et al.~\cite{Wang_2019_CVPR} first estimates an image-to-illumination mapping for modeling varying-lighting conditions and then takes this map to enhance underexposed images.
SID~\cite{Chen2018SID} converts raw short-exposure sensor images to RGB via a U-net-based denoising method followed by histogram stretching.
Many recent methods utilize the Retinex theory to design the CNN model.
E.g., LightenNet~\cite{li2018lightennet} and MSR-net~\cite{shen2017msr} are based on the single and multi-scale Retinex models respectively, to estimate the illumination map via CNNs.

To the best of our knowledge, none of the prior CNN methods leverage color features in dark images for LLIE.
Most CNN methods can enhance the image lightness to a great extent, however enhancing lightness alone is insufficient to generate high-quality RGB images with natural colors. 
Our proposed method introduces color attention maps from low-light images to be used as prior information for recovering natural high-quality images.

\Section{Proposed Method}

Our color-wise attention model decomposes the low-light RGB image, into lightness and color components via the LAB color space.
The motivation is to simplify the challenging LLIE process, and allow the color information drive the attention of CWAN$_{AB}$, while CWAN$_L$ focuses on enhancing image lightness and denoising simultaneously.
As in Fig.~\ref{fig:Motivation}, both the lightness and color components can be enhanced separately, and their fusion results in the final enhanced image.
Enhancing both of these components separately is, conceptually, easier compared to performing both tasks at the same time.

\SubSection{Problem formulation}

Given a low-light image, its lightness image $\mathbf{X}_L \! \in \! R^{H\times W}$ is fed to CWAN$_L$, and its color component images $\mathbf{X}_{AB} \! \in \! R^{H\times W \times2}$ are fed to CWAN$_{AB}$. 
CWAN$_L$ outputs the enhanced lightness image $\hat{\mathbf{X}}_L \! \in \! R^{H\times W}$.
CWAN$_{AB}$ outputs the enhanced color images $\hat{\mathbf{X}}_{AB} \! \in \! R^{H\times W \times2}$, along with two intermediate outputs, the color attention maps $\hat{\mathbf{M}} \! \in \! R^{H\times W \times2}$ and sparse attention points $\hat{\mathbf{P}} \! \in \! R^{H\times W \times2}$.
We aim to train CWAN$_L$, denoted by $\mathscr{F}_L(\mathbf{X}_L; \theta_L)$, to  map from low-light gray-scale image to an enhanced lightness image with reduced noise.
Similarly, we train CWAN$_{AB}$,  $\mathscr{F}_{AB}(\mathbf{X}_{AB}; \theta_{AB})$, to map from dark colors in low-light images to enhanced colors, under the constraint of color-wise attention. 
Thus, we formulate the LLIE problem as:
\vspace{-0.12 cm}
\begin{equation}
    \theta^*_L = \arg \min_{\theta_L} E_{\mathbf{X}_L,\mathbf{Y}_L,\mathscr{D}_L} [\mathscr{L}_L(\mathscr{F}_L), \mathbf{Y}_L ]  
\end{equation}
\vspace{-0.5 cm}
\begin{equation}
\label{eqn: AB}
    \theta^*_{AB} = \arg \min_{\theta_{AB}} E_{\mathbf{X}_{AB},\mathbf{Y}_{AB},\mathbf{P},\mathscr{D}_{AB}} [\mathscr{L}_{AB}(\mathscr{F}_{AB}), \mathbf{Y}_{AB},\mathbf{P} ], 
\end{equation}
where $\mathscr{D}$ denotes the training dataset, $\mathscr{L}$ denotes the loss function, $\mathbf{Y}_L  \! \in \! R^{H\times W}$ and $\mathbf{Y}_{AB}  \! \in \! R^{H\times W \times2}$ are the ground truth lightness and color components respectively, and $\mathbf{P}  \! \in \! R^{H\times W \times2}$ is a sparse set of ground truth attention points that are used to guide CWAN$_{AB}$.
As seen from Fig.~\ref{fig:Arch}, CWAN$_{AB}$ firstly estimates attention maps $\hat{\mathbf{M}}$ with the attention map generator $\mathscr{F}_{M}(\mathbf{X}_{AB}; \theta_{M})$ to help the learning of $\hat{\mathbf{P}}$. 
Before optimizing Eqn.~\ref{eqn: AB}, we pretrain $\mathscr{F}_{M}(\mathbf{X}_{AB}; \theta_{M})$ to generate $\hat{\mathbf{M}}$, which is formulated as:  
\begin{equation}
    \theta^*_M = \arg \min_{\theta_M} E_{\mathbf{X}_{AB},\mathbf{M},\mathscr{D}_{AB}} [\mathscr{L}_M(\mathscr{F}_M), \mathbf{M} ],
\end{equation}
where $\mathbf{M} \! \in \! R^{H\times W \times2}$ is the ground truth attention map. 
As described in Sec.~\ref{Sec: Attention map}, the attention maps and attention points serve different purposes. 
The former helps to identify local regions associated with foreground colors, while the latter specifies sparse points within the local region.

\SubSection{Network architecture}

CWAN utilizes two fully convolutional networks (FCN), such that both FCNs are composed of a feature extraction conv layer ($f_{ext}$), several convolutional blocks in the middle, and a final feature reconstruction conv layer ($f_{rec}$). 
The detailed structure is in Fig.~\ref{fig:Arch}.
We learn the residual using a global skip connection, rather than the direct mapping, to ease the training difficulty. 
All blocks have the same number of conv and ReLU layers.
Here, we define two types of blocks, memory blocks used in CWAN$_L$, and forward blocks used in CWAN$_{AB}$, as explained below.

\SubSubSection{CWAN$_L$ structure}
CWAN$_L$ is composed of a series of memory blocks.
These blocks are adopted from the image restoration work in~\cite{tai2017memnet}, which was successfully used for image denoising, super-resolution and JPEG deblocking.
We refer readers to~\cite{tai2017memnet}, for a more detailed explanation on memory blocks.
Generally, memory blocks utilize local short skip connections within the block to represent short-term memory, as well as long skip connections sourcing from previous blocks to represent long-term memory as in Fig.~\ref{fig:Arch}.
The short- and long-term memory help CWAN$_L$ to realize minor and major lightness enhancements within and across memory blocks.

\SubSubSection{CWAN$_{AB}$ structure}
\label{Sec: CWAN_AB}
In CWAN$_L$, all conv layers have the same number and size of filters, making long and short skip connection possible.
In contrast, CWAN$_{AB}$ does not utilize short and long skip connections.
Instead, in each block, the middle conv layer is a nonlinear activation with $1 \times 1$ filters.
This technique was successfully used in super-resolution~\cite{dong2016image}.

The CWAN$_{AB}$ network is composed of two parts.
The first part $\mathscr{F}_{M}(\mathbf{X}_{AB}; \theta_{M})$ takes $\mathbf{X}_{AB}$ as input to generate an attention map $\hat{\mathbf{M}}$; the second part takes both $\mathbf{X}_{AB}$ and $\hat{\mathbf{M}}$ forming a four-channel input to enhance colors.
The goal of the first part is to internally estimate $\hat{\mathbf{M}}$ with high activations at points of interest in an image, such that it guides the local regions during the enhancement of the second part.
Since $\mathbf{X}_{AB}$ has two color channels, the estimated $\hat{\mathbf{M}}$ is also a two-channel attention map, so that attention is for each channel at each spatial coordinate.
To supervise the learning of $\mathscr{F}_{M}$, we propose to use color frequency images to generate ground truth attention maps $\mathbf{M}$, described in Sec.~\ref{Sec: Attention map}. 
Therefore, the second part of CWAN$_{AB}$ learns the mapping from the stacked $\mathbf{X}_{AB}$ and $\hat{\mathbf{M}}$, to an enhanced $\hat{\mathbf{X}}_{AB}$ along with sparse attention color points $\hat{\mathbf{P}}$.
The ground truth attention points $\mathbf{P}$ are generated by selecting a set of non-zero foreground color points from $\mathbf{M}$.
Both $\mathbf{M}$ and $\mathbf{P}$ play major roles in our color-wise attention mechanism.

\begin{figure}[t!]
   \centering
\begin{tabular}{@{}c@{}c@{}c@{}c@{}}
\includegraphics[width=0.25\linewidth]{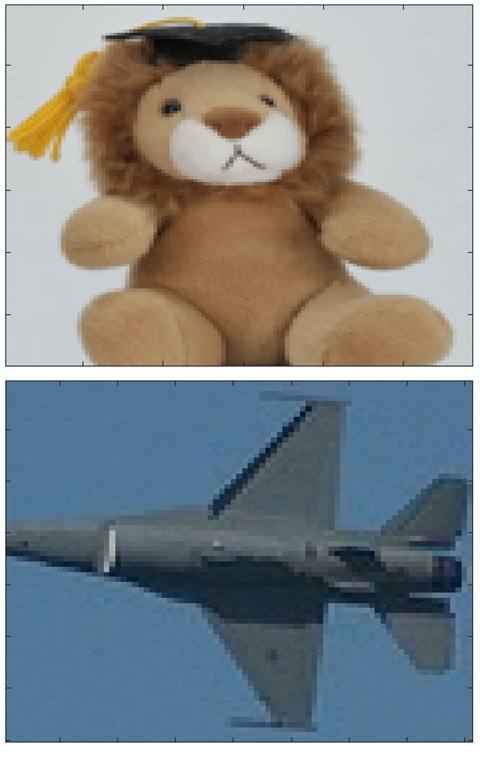}&
\includegraphics[width=0.25\linewidth]{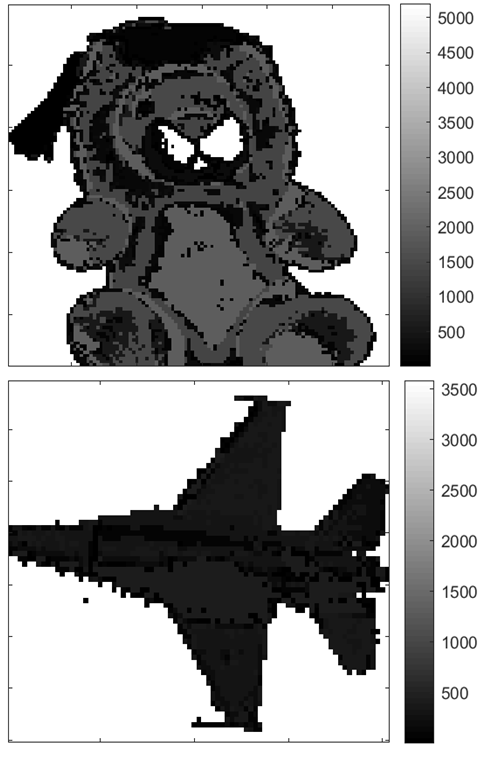}&
\includegraphics[width=0.25\linewidth]{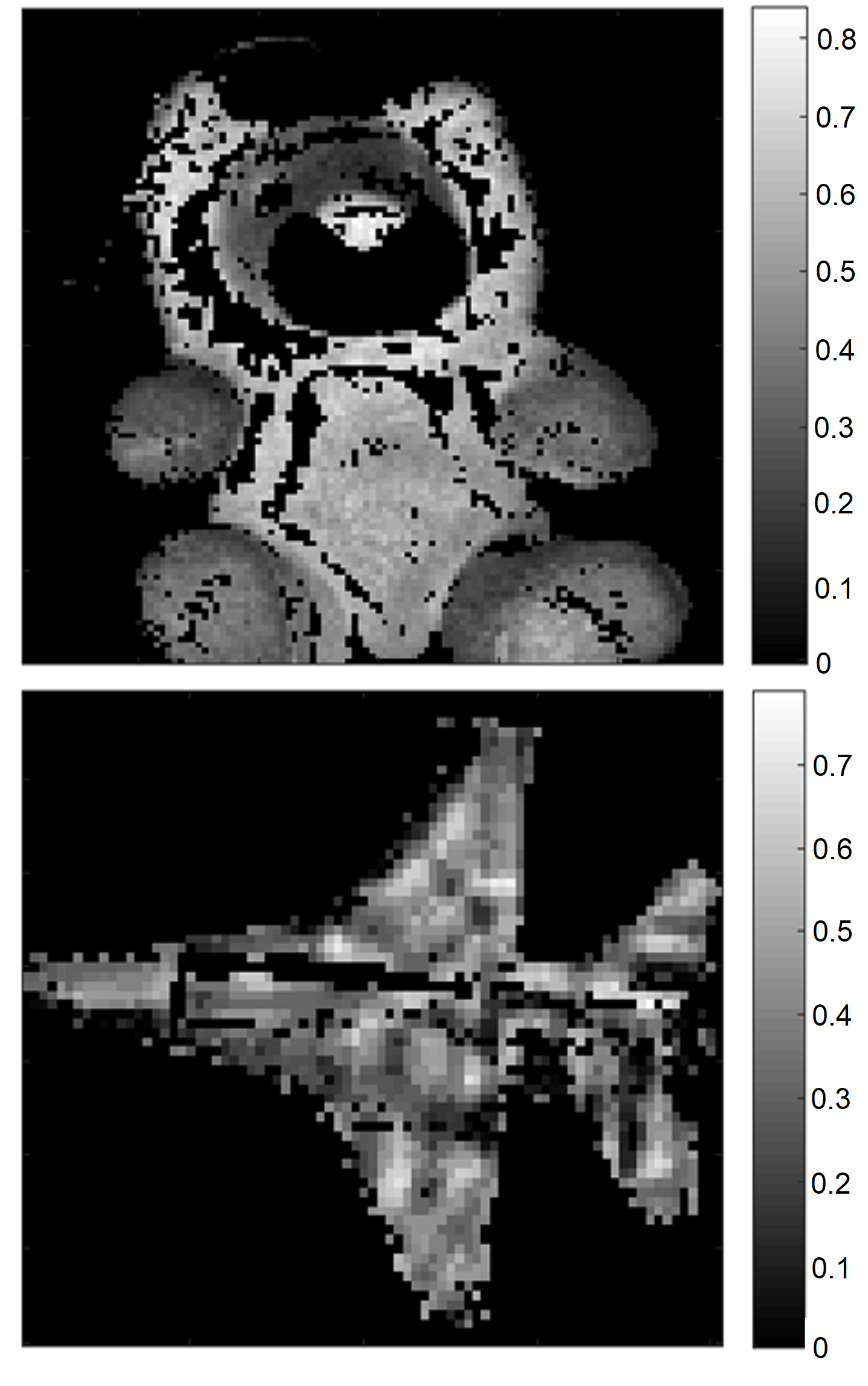}&
\includegraphics[width=0.25\linewidth]{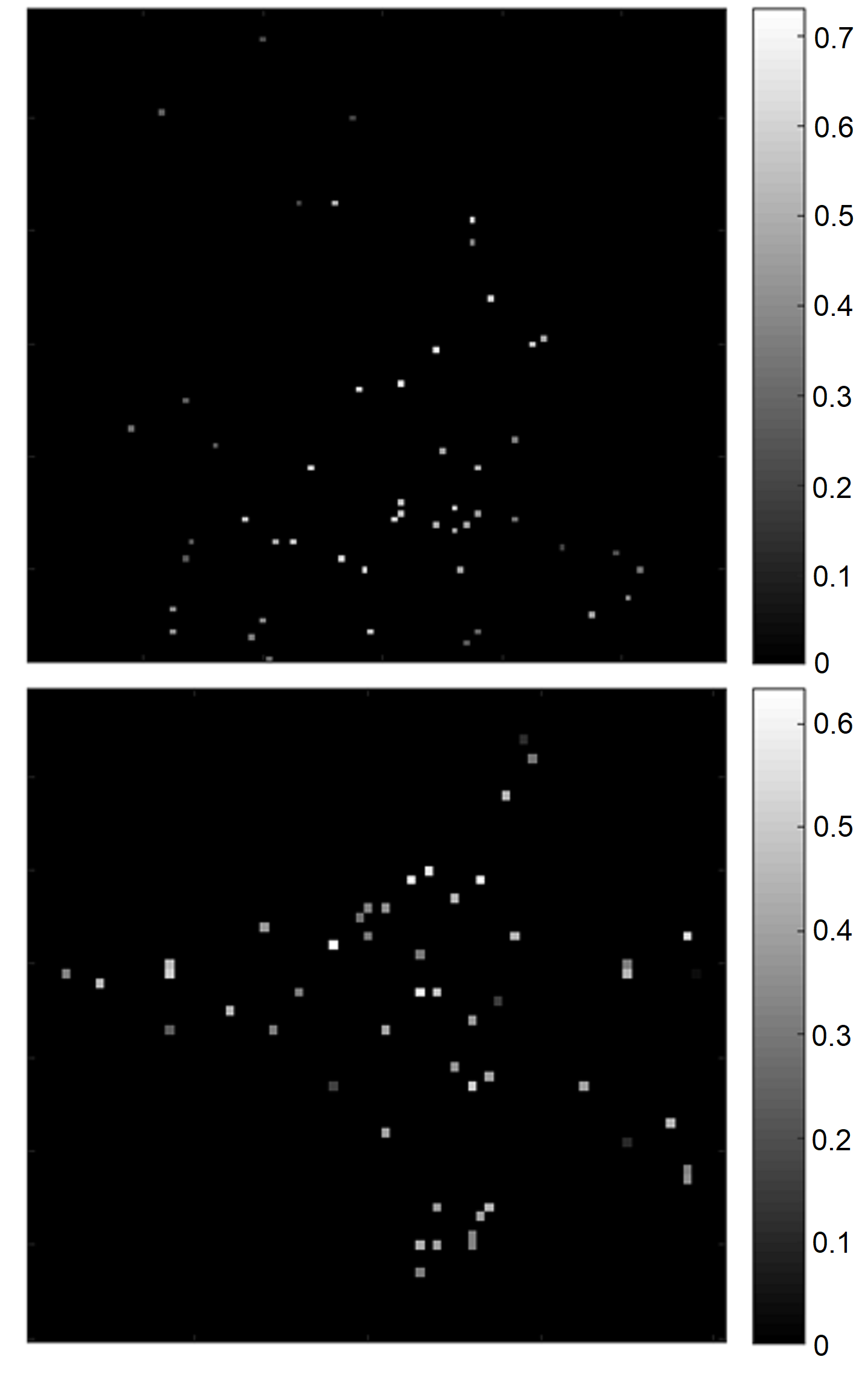}\vspace{-2mm}\\
\small{(a)}&\small{(b)}&\small{(c)}&\small{(d)}
\end{tabular}
    \caption{(a) input image, (b) color frequency image $\mathbf{F}$, (c) attention map $\mathbf{M}$, and (d) attention points $\mathbf{P}$. For $\mathbf{M}$ and $\mathbf{P}$, we show the first channel only. }
    \vspace{-0.2 cm}
    \label{fig:AttentionColor}
\end{figure}

\SubSection{Attention maps and points}
\label{Sec: Attention map}

A frequency image characterizes the spatial distribution along with the frequency information~\cite{kashiwagi2007introduction}. 
Given an image $\mathbf{X} \! \in \! R^{H\times W \times3}$, we can compute its {\it color frequency image} $\mathbf{F} \! \in \! R^{H\times W \times1}$, where $\mathbf{F}(x,y)$ equals to the number of occurrences of the RGB color $\mathbf{X}(x,y)$ in image $\mathbf{X}$.
Examples of color frequency images are in Fig.~\ref{fig:AttentionColor}.

We propose to utilize the color frequency image in CWAN$_{AB}$ as follows.
First, we apply a threshold $\tau$ on $\mathbf{F}$ to eliminate specific undesired frequencies.
E.g., the white background in the top image of Fig.~\ref{fig:AttentionColor} has very high frequency in $\mathbf{F}$.
Whereas some parts of $\mathbf{X}$ have very low frequencies in $\mathbf{F}$, e.g., noisy pixels or the eyes of the toy.
By segmenting $\mathbf{F}$ via $\tau_l < \mathbf{F} < \tau_u$, we emphasize our attention to foreground colors and eliminate both dominating color frequencies, and minor noisy regions. 
This results in a binary mask of desired color frequencies, denoted by $\bar{\mathbf{F}}$:
\begin{equation}
    \bar{\mathbf{F}}(x,y) =
    \begin{cases}
        1 , &  \text{if} \,\,\, \tau_l < \mathbf{F}(x,y) < \tau_u \\
        0  , & \text{otherwise}.
    \end{cases}
\end{equation}
After computing $\bar{\mathbf{F}}$, we generate the ground truth color attention map $\mathbf{M}(:,:,i) = \mathbf{X}_{AB}(:,:,i) \odot \bar{\mathbf{F}}$, where $\odot$ is a Hadamard product, and $i \in [1,2]$. 
We apply linear normalization on $\mathbf{M}$ such that it is in the range of $[0,1]$. 
Usually $\mathbf{M}$ contains foreground colors as seen in Fig.~\ref{fig:AttentionColor}.

In~\cite{zhang2017real}, authors learn a deep network for gray-scale image colorization, where the user can interact and guide the colorization process via {\it manually} selecting colors of specific pixels.
Differs to the manual selection, CWAN$_{AB}$ {\it randomly} selects a finite set of non-zero foreground color points $\beta$ from $\mathbf{M}$ to guide the color-wise attention model.
We define a binary mask $\mathbf{B}_P$ assigned with $1$'s at the coordinates of all $\beta$ points, such that $\sum\sum \mathbf{B}_P = \beta$.
Then, we compute the ground truth attention points $\mathbf{P}$ to represent a sparse subset of foreground color points, via $\mathbf{P}(:,:,i) =  \mathbf{M}(:,:,i)  \odot \mathbf{B}_P$, where $i \in [1,2]$. 

In our attention mechanism,  $\mathbf{M}$ supervised the learning of $\hat{\mathbf{M}}$, which  inputs to  the subsequent network and guides CWAN$_{AB}$ to focus on enhancing foreground colors at {\it coarse local regions}.
On the other hand, $\mathbf{P}$ guides the color enhancement at {\it a pixel level}, identifying key foreground  colors. 
Since $\mathbf{M}$ has much more duplicated foreground colors than $\mathbf{P}$, supervising by $\mathbf{P}$ can cover majority of colors, yet using minimal, not duplicated, constraints.


\SubSection{Objective function}
\label{Loss}
To train CWAN$_L$, we use the $L_1$ loss function to treat synthesizing the enhanced lightness $\hat{\mathbf{X}}_L$ as a regression problem.
On the other hand, training CWAN$_{AB}$ takes place in two stages.
The first stage trains the color-wise attention map generator to predict $\hat{\mathbf{M}}$ only, via the $L_1$ loss:
\begin{equation}
\label{EQ: M_Loss}
    \mathscr{L}_{M}(\mathscr{F}_{M}(\mathbf{X}_{AB}; \theta_{M}), \mathbf{M}) =  ||\hat{\mathbf{M}} - \mathbf{M}||_1.   
\end{equation}

The second stage learns CWAN$_{AB}$ end-to-end, including fine tuning the pretrained attention map generator.
We propose to use the following loss function:
\begin{equation}
\label{EQ: AB_Loss}
    \mathscr{L}_{AB}(\mathscr{F}_{AB}(\mathbf{X}_{AB} ; \theta_{AB}), \mathbf{Y}_{AB}) = \mathscr{L_H} + \alpha \mathscr{L}_{MSE}.
\end{equation}
Here $\alpha$ is the weight, $\mathscr{L_H}$ is the Huber loss applied to $\hat{\mathbf{X}}_{AB}$.
The Huber loss has witnessed great success in the image colorization field~\cite{Yoo_2019_CVPR,zhang2017real}, due to  the relative high color saturation effect, making it suitable for enhancing low-light images.
Further, we choose the Huber loss also because it is a robust estimator, and can help to avoid the averaging problem. 
The $\mathscr{L}_{MSE}$ is the mean square error loss applied to the estimated attention points $\hat{\mathbf{P}}$, as follows:
\begin{equation}
\label{EQ: HUBER}
    \mathscr{L_H} =
    \begin{cases}
        \frac{1}{2}( \hat{\mathbf{X}}_{AB} - \mathbf{Y}_{AB} )^2 , &  \text{if} \,\, |\hat{\mathbf{X}}_{AB} - \mathbf{Y}_{AB} | \leq \delta \\
        \delta | \hat{\mathbf{X}}_{AB} - \mathbf{Y}_{AB} | -  \frac{1}{2}\delta^2  , &  \text{otherwise}
    \end{cases}
\end{equation}
\begin{equation}
\label{EQ: lmse}
   \mathscr{L}_{MSE} = \frac{1}{\beta}|| (\hat{\mathbf{P}} - \mathbf{P})\odot \mathbf{B}_P ||_2^2, 
\end{equation}
where $\delta$ is the parameter of the Huber loss.
While CWAN$_{AB}$ outputs colors at all locations in $\hat{\mathbf{P}}$, $\mathscr{L}_{MSE}$ loss is computed only using reconstructed colors at randomly sampled color locations in $\mathbf{B}_P$.

\Section{Experiments}

This section provides ablation, quantitative and qualitative results, on both real-world and synthetic low-light images. 
We use metrics including Structural Similarity (SSIM), Peak Signal-to-Noise Ratio (PSNR),  Lightness-Order-Error (LOE), Colorfulness (C), and a case study.

\Paragraph{Datasets}
We use the See-In-the-Dark (SID) database~\cite{Chen2018SID} to learn CWAN, containing $5,094$ raw low-light images, and $424$ RGB ground truth images, such that multiple low-light images correspond to the same ground truth image. 
The data is divided into two distinct subsets, one captured with a Sony camera (SID$_{Sony}$), and another with a Fuji camera (SID$_{Fuji}$). 
We follow the same protocol in~\cite{Chen2018SID} to divide data into training, validating and testing sets. 
Since most low-light literature uses RGB as input, we convert the low-light images from raw to RGB using a COTS converter (easy2convert.com) which supports raw images of both Sony and Fuji cameras.
The generated RGB images contain less noise than the original SID dataset, however, the low-light condition remains severe.

We further generate a synthetic low-light dataset from PASCAL VOC~\cite{pascal-voc-2012}, which contains realistic scenes of objects.
With randomly selected $1,000$ images, we synthesize low-light images following a Retinex-based approach~\cite{li2018lightennet}, with a $85\%$ decrease in the pixel intensity.
We refer to this as PASCAL$_{1000}$ dataset.
Finally, we evaluate CWAN using low-light images collected by~\cite{EKDMU17HDR} named HDRDB, which does not have groundtruth.
With $96$ low-light images of natural scenes, HDRDB has low-exposure, Gaussian noise, $5\%$ of pixels are saturated and contain no information, with an additional $60\%$ decrease on the pixel intensity using~\cite{li2018lightennet}.

\Paragraph{Experimental parameters}
For CWAN$_L$, we use three memory blocks, i.e., total of $11$ conv layers, with the weight decay of $0.05$ and batch size of $16$. 
The large weight decay as in~\cite{dong2016image}, helps to improve generalization on unseen low-light images.
The number of patches from SID$_{Sony}$ is $37,300$ and from SID$_{Fuji}$ is $33,100$, of size $64\! \times \!64$, i.e., $50$ patches per training image.
For CWAN$_{AB}$, we use one forward block in the attention map generator, i.e., total of $5$ conv layers, and two forward blocks for the color enhancement part, i.e., total of $8$ conv layers, with the weight decay of $0.05$ and batch size of $32$.
The number of patches in training CWAN$_{AB}$ from SID$_{Sony}$ is $186,500$ and from SID$_{Fuji}$ is $165,500$, of size $32\! \times\! 32$, i.e., $100$ patches per training image. 
The patch sizes ($32$ or $64$) were experimentally determined. 
We use $\tau_l = 0.05N$ and $\tau_u = 0.5N$ during training on small patches, where $N$ is the number of pixels in the image.
Other parameters in CWAN$_{AB}$ are $\beta = 20$, $\alpha=1$ and $\delta= 0.5$.
The learning rate for all CNNs is $10^{-4}$, and trained for $200$ epochs.
We implement CWAN on NVIDIA GTX1080Ti GPU, with Matconvnet toolbox~\cite{vedaldi2015matconvnet}.

\SubSection{Ablation Study}
\begin{table}[t!]
\centering
\resizebox{6.5cm}{!}{
\begin{tabular}{ |c|c|c|c| } 
\hline
Test image & HE & LLNet~\cite{lore2017llnet}  & CWAN$_L$ \\
\hline
Bird-D & $11.28$ / $0.62$  & $18.43$ / $0.60$  &  $\mathbf{28.76}$ / $ \mathbf{0.91}$ \\ 
Bird-D+GN & $9.25$ / $0.09$  & $19.73$ / $0.56$  &  $\mathbf{24.21}$ / $ \mathbf{0.79}$ \\ 
\hline
Girl-D & $18.27$ / $0.80$  &  $22.45$ / $0.80$ &  $\mathbf{24.90}$/ $ \mathbf{0.83}$ \\ 
Girl-D+GN & $16.07$ / $0.26$  &  $20.04$ / $0.60$ &  $\mathbf{21.96}$/ $ \mathbf{0.77}$ \\ 
\hline
House-D &  $12.03$ / $0.65$ & $21.10$ / $0.64$  & $\mathbf{21.55}$ / $\mathbf{0.70}$ \\
House-D+GN &  $10.55$ / $0.33$ & $\mathbf{20.25}$ / $0.56$  & $19.98$ / $\mathbf{0.62}$ \\
\hline
Pepper-D & $18.45$ / $\mathbf{0.85}$ & $21.33$ / $0.78$ &  $\mathbf{25.48}$ /  $0.83$ \\
Pepper-D+GN & $14.69$ / $0.21$ & $22.33$ / $0.78$ &  $\mathbf{24.25}$ /  $\mathbf{0.80}$ \\
\hline
Town-D & $17.55$ / $0.79$ &  $22.47$ / $0.81$ & $\mathbf{26.62}$ /  $\mathbf{0.87}$ \\
Town-D+GN & $14.85$ / $0.25$ &  $20.00$ / $0.60$ & $\mathbf{23.11}$ /  $\mathbf{0.79}$ \\
\hline
\end{tabular}
}
\caption{Comparing lightness estimation in PSNR/SSIM. Here, D refers to a dark image generated using~\cite{lore2017llnet}, and GN refers to Gaussian noise with $\sigma = 18$.}
\label{TABLE: Gray_baseline}\figvspace
\end{table}
\Paragraph{Analysis of CWAN$_{L}$}
We compare the performance of CWAN$_{L}$ with LLNet~\cite{lore2017llnet} which uses a stacked auto-encoder for joint low-light enhancement and denoising, trained using gray-scale images~\footnote{Dataset URL: http://decsai.ugr.es/cvg/dbimagenes/}, i.e., lightness enhancement only.
We fine-tune CWAN$_L$ using the same train set, on only $1,600$ patches of size $64\! \times\! 64$ for two epochs.
We follow the same low-light test set generation as in~\cite{lore2017llnet}, which only consists of $5$ images.
The results are shown in Tab.~\ref{TABLE: Gray_baseline}, and visual results are in the supplementary file.
CWAN$_L$ on average is able to outperform HE and LLNet on all five images, demonstrating the effectiveness of lightness enhancement and denoising in CWAN.
Nevertheless, the huge PSNR/SSIM margin in Tab.~\ref{TABLE: Gray_baseline} is mainly due to differences in the network structures and training datasets.

\Paragraph{Loss function weights of CWAN$_{AB}$}
To understand the benefit of the attention model, we ablate the weights in the CWAN$_{AB}$ loss function in Eqn.~\ref{EQ: AB_Loss}.
We first train CWAN$_{AB}$ with equal weights, $\alpha = 1$.
Starting with this model, we fine-tune the network for $5$ epochs independently with different $\alpha$ between $1.4$ down to $0$.
This experiment is conducted on the train/test sets of SID$_{Sony}$. 
As in Tab.~\ref{TABLE: ablation1}, our model heavily relies on the attention mechanism in low-light image enhancement.
When the weights of both terms are equal, we reach a top PSNR of $28.56$.
When $\alpha > 1$, i.e., the attention points have more impact than the Huber loss in training CWAN$_{AB}$, the PSNR drops significantly. 
On the other hand, when $\alpha = 0$, i.e., CWAN$_{AB}$ is trained without the attention mechanism, the PSNR drops to $26.31$. 

\begin{table}[t!]
\small
\centering
\resizebox{8.4cm}{!}{
\begin{tabular}{ |c|c|c|c|c|c|c|c|c| } 
\hline
$\alpha$ & $1.4$ & $1.2$ & $1.0$ & $0.8$& $0.6$& $0.4$& $0.2$& $0.0$ \\
\hline
PSNR  & $26.46$ & $27.88$  & $\mathbf{28.56}$ & $28.35$& $27.98$& $27.48$& $27.32$& $26.31$ \\ 
\hline
\end{tabular}
}
\caption{Loss function weight analysis on CWAN$_{AB}$.}
\label{TABLE: ablation1}\figvspace
\end{table}

\Paragraph{Memory vs.~forward blocks}
We adopt the memory and forward block concepts from low-level vision literature ~\cite{tai2017memnet,dong2016image}, and incorporate them in CWAN$_L$ and CWAN$_{AB}$ respectively.
We ablate all possible combinations of using two types of blocks in our model, and the PSNRs are as follows given the order of (CWAN$_L$/CWAN$_{AB}$): $27.23$ with (forward/forward), $27.75$ with (forward/memory), $28.56$ with (memory/forward) and $28.17$ with (memory/memory).
Both types of blocks perform well in our CWAN, with the (memory/forward) configuration of (CWAN$_L$/CWAN$_{AB}$) being the best.
Our method makes it possible to incorporate future novel block designs into our network.

\Paragraph{Analysis on decomposing image into L and AB }
The intuitive alternative to our decomposition technique, is to learn a direct mapping from the input low-light color image to an enhanced RGB image. 
We train two baseline FCN models, using RGB and LAB input data named FCN$_{RGB}$ and FCN$_{LAB}$ respectively.
Note that FCN$_{LAB}$ estimates a LAB image, which is then converted to RGB.
Both networks have the exact same structure as CWAN$_{L}$, with the same hyper-parameters. 
Using the test set of SID$_{Sony}$, FCN$_{RGB}$ and FCN$_{LAB}$ achieve a PSNR of $25.71$ and $22.58$ respectively, whereas CWAN can reach a PSNR of $28.56$.
This shows how effective our decomposition technique over the traditional feed-forward FCN approach. 

\Paragraph{Alternatives of CWAN$_{AB}$} 
We introduce alternative methods of CWAN$_{AB}$ for comparison using the SID$_{Sony}$ dataset. 
(1) CWAN$^C_{AB}$ removes the attention map generator, and only inputs $\mathbf{X}_{AB}$ to CWAN$_{AB}$.
(2) CWAN$^M_{AB}$ inputs the attention map $\mathbf{M}$, computed by the color frequency image $\mathbf{F}$ without the attention map generator, along with $\mathbf{X}_{AB}$ to CWAN$_{AB}$.  
(3) CWAN$^P_{AB}$ inputs the attention points $\mathbf{P}$ along with $\mathbf{X}_{AB}$.
All three architectures are trained on SID$_{Sony}$, while using the same CWAN$_L$ for lightness enhancement.
As in Tab.~\ref{TABLE: T1}, CWAN$^C_{AB}$ performs poorly without the attention map, but achieves higher PSNR than FCN$_{LAB}$ which proves the usefulness of image decomposition.
CWAN$^M_{AB}$ achieves the closest result to CWAN$_{AB}$ with a significant margin of $0.63$ in PSNR.
This shows that when training CWAN$_{AB}$ end-to-end, the attention map generator is able to learn a more effective $\hat{\mathbf{M}}$ than $\mathbf{M}$.
Thus, despite we could directly compute $\mathbf{M}$ from $\mathbf{X}_{AB}$, it is better to let the attention map generator produce a better map $\hat{\mathbf{M}}$ for subsequent enhancement.
CWAN$^P_{AB}$ proves that having a finite set of points guiding the network attention can achieve high PSNR, but not higher than feeding $\mathbf{M}$.

\begin{table}[t!]
\small
\centering
\resizebox{7.4cm}{!}{
\begin{tabular}{ |c|c|c|c|c| } 
\hline
Architectures  & CWAN$^C_{AB}$ &  CWAN$^M_{AB}$ & CWAN$^P_{AB}$ &  CWAN$_{AB}$ \\
\hline
Input of CNN & $\mathbf{X}_{AB}$ & $\mathbf{X}_{AB}$ \& $\mathbf{M}$ & $\mathbf{X}_{AB}$ \& $\mathbf{P}$ & $\mathbf{X}_{AB}$ \\
Estimates $\hat{\mathbf{M}}$  & \xmark  & \xmark & \xmark  & \checkmark  \\
Estimates $\hat{\mathbf{P}}$ & \xmark  & \checkmark  & \checkmark  & \checkmark  \\
Loss function & Eqn.~\ref{EQ: HUBER} & Eqn.~\ref{EQ: AB_Loss}  & Eqn.~\ref{EQ: AB_Loss}   & Eqn.~\ref{EQ: AB_Loss}   \\
PSNR & $25.86$ & $27.93$ & $27.39$ & $\mathbf{28.56}$  \\
SSIM &  $0.859$ & $0.890$ & $0.887$ & $\mathbf{0.909}$ \\
\hline
\end{tabular}
}
\caption{Compare CWAN$_{AB}$ with alternative architectures.} 
\label{TABLE: T1} \vspace{-0.3 cm}
\end{table}

\begin{figure}[t!]
\begin{center}
   \includegraphics[width=0.95\linewidth]{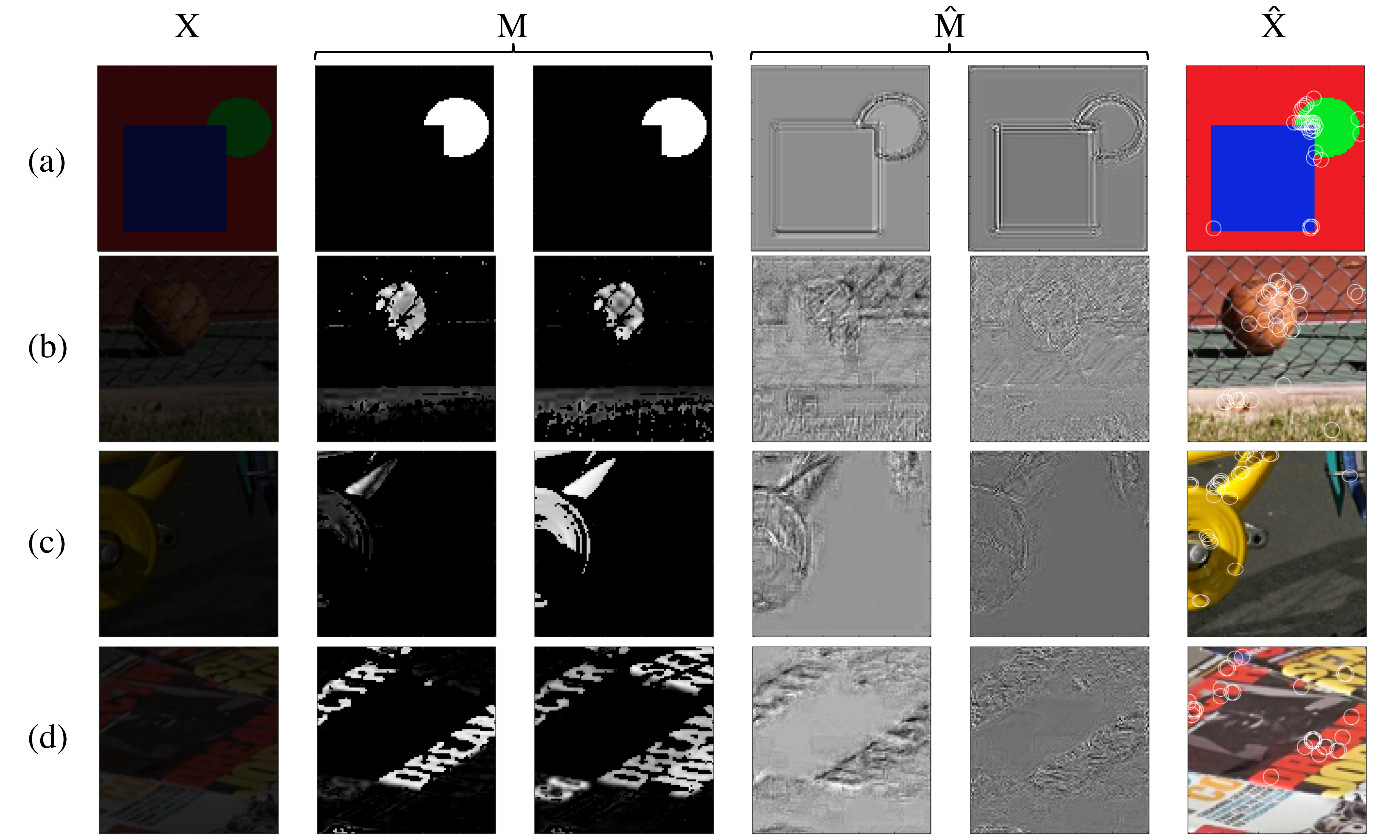}
\end{center}
\figvspace
   \caption{$\mathbf{M}$ vs. $\hat{\mathbf{M}}$: (a) a toy example, (b-d) low light images. The four middle columns are normalized in the range of $[0,1]$, and represent the two channels of both $\mathbf{M}$ and $\hat{\mathbf{M}}$ respectively. the highest $30$ responses in $\hat{\mathbf{M}}$ are overlaid in $\hat{\mathbf{X}}$ in the right column. }
\label{fig:ablation3} \figvspace
\end{figure}

\Paragraph{Attention map analysis}
The attention map generator aims to select pixels with desired color frequencies. 
As in Fig.~\ref{fig:ablation3}, our attention map generator is trained with $\mathbf{M}$ to generate $\hat{\mathbf{M}}$ highlighting foreground colors.
After training CWAN$_{AB}$ end-to-end, the conv layers of the attention map generator were fine-tuned to estimates $\hat{\mathbf{M}}$ with high responses at colors of interest in the scene; it also tends to produce higher responses near object edges where gradients exist, as in (a).
Note $\hat{\mathbf{M}}$ has no high responses in the background, as it occupies a large percentage of image, e.g., the street in (c). 
On the other hand, image (b) has various colors in the scene and $\hat{\mathbf{M}}$ produces high responses throughout the map, but tends to give the ball higher attention.

\Paragraph{Attention point analysis} 
$\mathbf{P}$ is constructed by selecting $\beta$ random points from $\mathbf{M}$. 
We study the impact of randomness in point selection, by independently fine-tuning CWAN$_{AB}$ three times, where each time we regenerate the $\beta$ points and train for five epochs.
By testing the three models on the test set, the standard deviation of the PSNR is merely $0.002$.
This desired low impact is due to the controlled randomness, where we only select points from $\mathbf{M}$, but not $\mathbf{X}_{AB}$. 

\begin{figure}[t]
\begin{center}
   \includegraphics[width=0.63\linewidth]{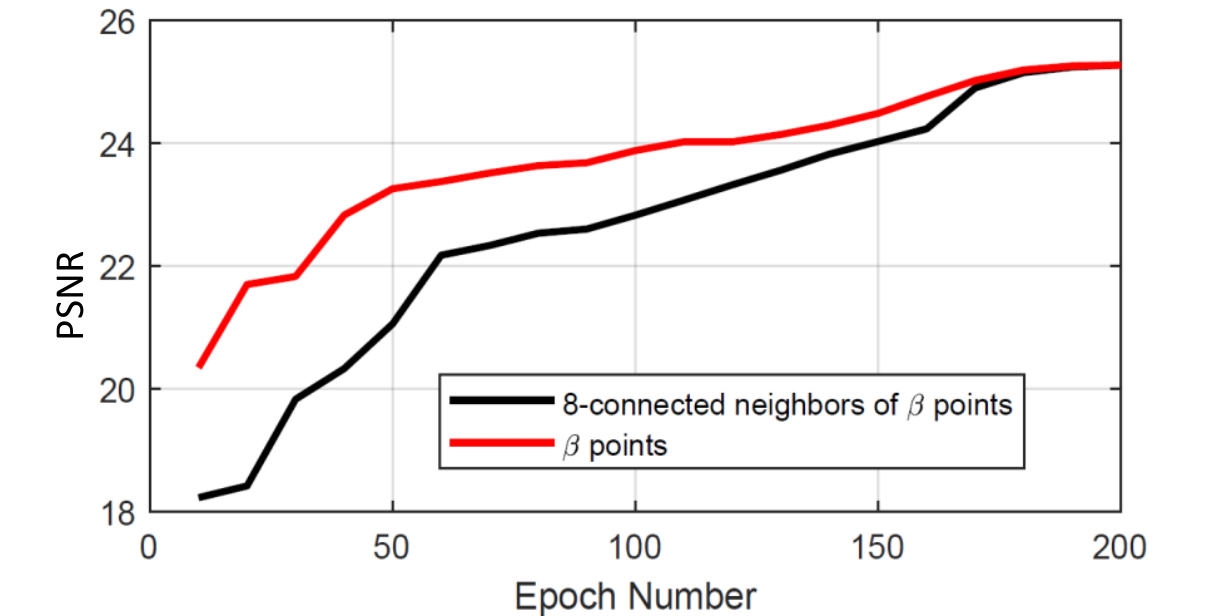}
\end{center}
\vspace{-0.4 cm}
   \caption{Color attention mechanism through the selected $\beta$ points from ${\mathbf{P}}$ in guiding the neighboring pixels.}
   \vspace{-0.2 cm}
\label{fig:Epoch_PSNR} \figvspace
\end{figure}

\Paragraph{How does the color attention mechanism work?}
The random points selected in ${\mathbf{P}}$, while training CWAN$_{AB}$ via the loss in Eqn.~\ref{EQ: lmse}, has a major impact in reconstructing the colors in $\hat{\mathbf{X}}_{AB}$. 
To demonstrate the effectiveness of this attention mechanism, 
we study the effect of enhancing the colors at the $\beta$ points compared to the local region surrounding those points.
While training CWAN$_{AB}$ end-to-end for various epochs, we compute the PSNR of only the $\beta$ selected points  as in Fig.~\ref{fig:Epoch_PSNR}, i.e., excluding all other pixels in the image.
We compare the result with the PSNR of the $8$-connected neighboring pixels surrounding the $\beta$ points, i.e., $3\!\times\!3$ local regions excluding the center $\beta$ points.
Fig.~\ref{fig:Epoch_PSNR} illustrates how CWAN$_{AB}$ learns to enhance the colors at $\beta$ points promptly, and gradually pulls the neighboring pixels for color enhancement. 

One potential concern resides in the possibility of ignoring some colors in an image given the small value of $\beta$.
To study how CWAN$_{AB}$ selects a diverse set of colors, we first divide the colorful AB space into $40$ clusters each with a unique color.
Then we select a set of $2,000$ training patches $\mathbf{X}_{AB}$ along with their ground truth $\mathbf{Y}_{AB}$.
For every $\mathbf{Y}_{AB}$, we assign all pixels to the corresponding cluster via nearest neighboring, forming a color map $\mathbf{C}$. 
For every $\mathbf{X}_{AB}$, we allow CWAN$_{AB}$ to select $\beta=20$ points at locations defined by $\mathbf{B}_P$. 
Based on the selected points, we evaluate the color clusters that have been selected from the non-zero elements of $\mathbf{C} \odot \mathbf{B}_P$, and compare them with all available colors in $\mathbf{C}$.
By applying cross validation on three separate sets, the average percentage of selecting all possible colors is $88.3\%\pm1.3\%$.
This means that $\sim 90\%$ colors are selected by CWAN$_{AB}$, and the low standard deviation resembles consistency among the different sets.
This study shows one strength of our color attention mechanism.
That is, despite using a very small number of constraints, i.e., $\beta=20$ points in $\mathscr{L}_{MSE}$, our optimization has an impact on the majority ($90\%$) of to-be-recovered colors.

\begin{table*}[t!]
\centering
\resizebox{13cm}{!}{
\begin{tabular}{ |c|c|c|c||c|c|c|c| } 
\hline
\multirow{2}{*}{Method (training set)} & \multicolumn{3}{|c||}{PSNR / SSIM} & LOE &  Colorfulness  & AVG rank & \multirow{2}{*}{LLIE type}\\ 
\cline{2-7}
 & SID$_{Sony}$ & SID$_{Fuji}$ &  PASCAL$_{1000}$ & \multicolumn{3}{|c|}{HDRDB}  & \\ 
\hline
HE                          & $21.68/0.712$         & $20.25/0.726$       & $26.28 /  0.885$ & $46.53$ &   $14.87$ & $5.49\pm4.19$ & Generic\\ 
Dong~\cite{Dong2011Fast}    & $20.56/0.824$         & $22.40/0.841$       & $20.79 / 0.907 $ &  $78.88$  & $28.16$ &  $9.60\pm3.55$ & Generic\\ 
BIMEF~\cite{ying2017bio}    & $ 15.55 / 0.697 $     & $16.52/0.708$       & $  14.13 /0.732 $ & $38.78$  & $28.23$ & $9.94\pm1.79$ & Generic\\
Ying~\cite{Ying2017Neww}    & $18.61/0.746$         & $20.12/0.751$       & $ 18.24 / 0.810 $ & $36.25$ & $13.06$ & $9.58\pm1.98$ & Generic\\ 
JED~\cite{ren2018joint}     & $15.05/0.712$         & $15.96/0.716$       & $13.49 / 0.685 $  & $53.37$  & $13.75$ & $8.21\pm1.95$& Generic\\ 
AMSR~\cite{lee2013adaptive} & $12.86/0.567$         & $12.31/0.414$       & $ 12.38 / 0.626 $  &$318.83$ & $16.29$ & $10.21\pm2.12$ & Generic\\
LIME~\cite{guo2017lime}     & $19.10/0.802$         & $21.39/0.830$       & $ 20.72 /  \textcolor{red}{0.935} $ & $105.23$  & $\textcolor{blue}{32.02}$ &  $7.28\pm4.12$ & Generic\\ 
Li~\cite{li2018structure}   & $15.17/0.714$         & $15.96/0.711$       & $13.49/0.690$   & $51.95$  & $19.46$ & $9.60\pm2.38$ & Generic\\ 
LECARM~\cite{ren2018lecarm}   & $17.99/0.831$         & $18.39/0.829$       & $16.86/0.884$   & $\textcolor{blue}{32.83}$  & $19.17$ & $5.16\pm1.84$ & Generic\\
LightenNet~\cite{li2018lightennet}& $19.21/0.817$   & $17.87/0.812$       & $ 17.82 / 0.858 $ & $257.04$  & $28.76$ & $\textcolor{blue}{3.07\pm2.05}$  & CNN\\
MBLLEN~\cite{Lv2018MBLLEN}& $14.99/0.687$   & $13.53/0.621$       & $17.36/0.778$ & $39.03$  & $16.80$ & $7.18\pm2.18$  & CNN\\ 
SID~\cite{Chen2018SID} (SID$_{Sony}$)& \textcolor{blue}{$27.42$}$/0.877$  & $26.71/0.880$ & $ 24.91/0.872 $ & $35.40$ & $26.56$ & $3.37\pm1.43$ & CNN\\ 
\hline 
CWAN (SID$_{Sony}$)       &  \textcolor{red}{$28.56/0.909$}&  \textcolor{red}{$28.11/0.911$} & $  \textcolor{red}{29.08} /  \textcolor{blue}{0.924} $ &  $ \textcolor{red}{30.53}$  & $\textcolor{red}{40.38}$ & $\textcolor{red}{2.32\pm1.73}$ & CNN\\ 
CWAN (SID$_{Fuji}$)       &  $27.28/0.902$&  \textcolor{blue}{$26.77/0.911$}  & $ \textcolor{blue}{28.68}/0.923$ & ---& ---& --- & CNN \\
\hline
\end{tabular}
}
\caption{Quantitative results on four datasets. \textcolor{red}{Red}/\textcolor{blue}{blue} fonts indicate the best/second best results.}
\label{TABLE: Quantitative}
\end{table*}

\begin{figure*}[t]
\begin{center}
   \includegraphics[width=0.92\linewidth]{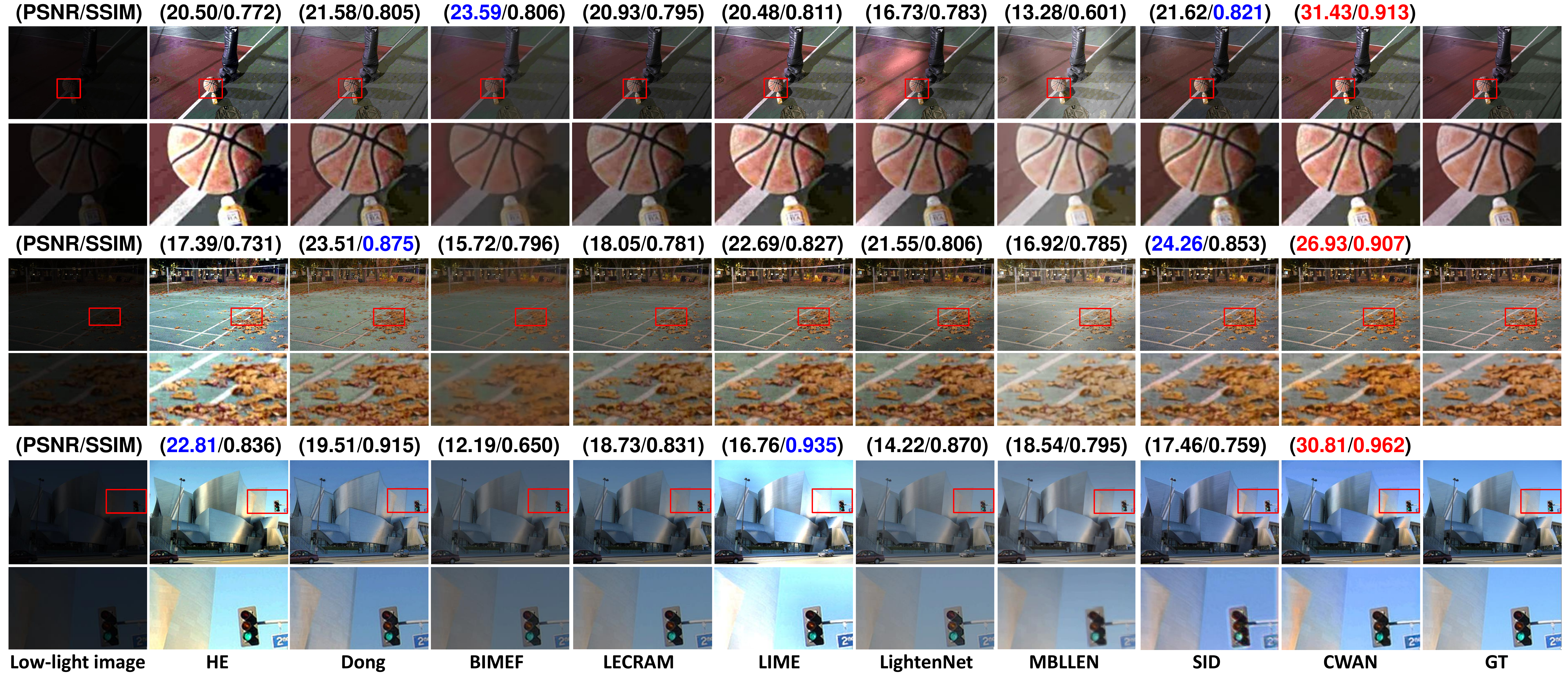}
\end{center}
\vspace{-0.4 cm}
   \caption{Qualitative results of various methods compared to CWAN on SID$_{Sony}$ (top), SID$_{Fuji}$ (middle) and PASCAL$_{1000}$ (bottom). }
\label{fig:Qualitative_pascal}\figvspace
\end{figure*}

\SubSection{Quantitative results}

We evaluate our method quantitatively on SID$_{Sony}$, SID$_{Fuji}$, PASCAL$_{1000}$, and HDRDB, comparing with $12$ methods, including both generic and CNN-based LLIE methods. 
For comparison, we use the published codes for the first $11$ methods, where SID is reimplemented by training a U-net structure with RGB data from SID$_{Sony}$ via a $L_1$ loss.
Tab.~\ref{TABLE: Quantitative} summarizes the results. 

The PSNR and SSIM are commonly used in LLIE literature to test image similarity with ground truth.
Our method achieves the best result on all datasets except for the SSIM of PASCAL$_{1000}$, where LIME has the highest value.
The LOE metric is commonly used for real low-light images~\cite{ying2017bio,Ying2017Neww,guo2017lime}.
Smaller LOE value means better lightness order is preserved, i.e., the intensity order of each pixel with all other pixels is similar between low-light and enhanced images. 
In Tab.~\ref{TABLE: Quantitative}, CWAN has the smallest LOE among all methods, demonstrating how well CWAN enhances low-light images while preserving the lightness order.
We further utilize the colorfulness metric~\cite{hasler2003measuring}, which estimates the quality of colors in an image.
Here, colorfulness represents the intensity and assortment of colors in the image, defined as $\sqrt{\sigma_{C_1}^2+\sigma_{C_2}^2} + 0.37\sqrt{\mu_{C_1}^2 + \mu_{C_2}^2}$, where $\mu$ and $\sigma$ are the mean and standard deviation of $C_1 = R - G$ and $C_2 = 0.5(R + G) - B$.
Tab.~\ref{TABLE: Quantitative} concludes that CWAN can synthesize more colorful images than all other methods.
This proves how effective the proposed color-wise attention technique in faithfully restoring the color of the scene.   

Finally, we conduct a user study on $20$ low-light images in HDRDB which includes several indoor and outdoor scenes.
We enhance the images with $12$ baselines and CWAN resulting in a total of $13$ images per low-light image.
$10$ people were asked to rank the images from the best (rank $1$) to the worst image (rank $13$).
We summarize the scores in Tab.~\ref{TABLE: Quantitative}.
The user study results demonstrate that CWAN has the best average rank of $2.32$, and that it can generate visually appealing images, with great color attributes. 
CWAN also has a small standard deviation indicating the high consensus among $10$ people.

\begin{figure*}[t!]
\begin{center}
   \includegraphics[width=0.91\linewidth]{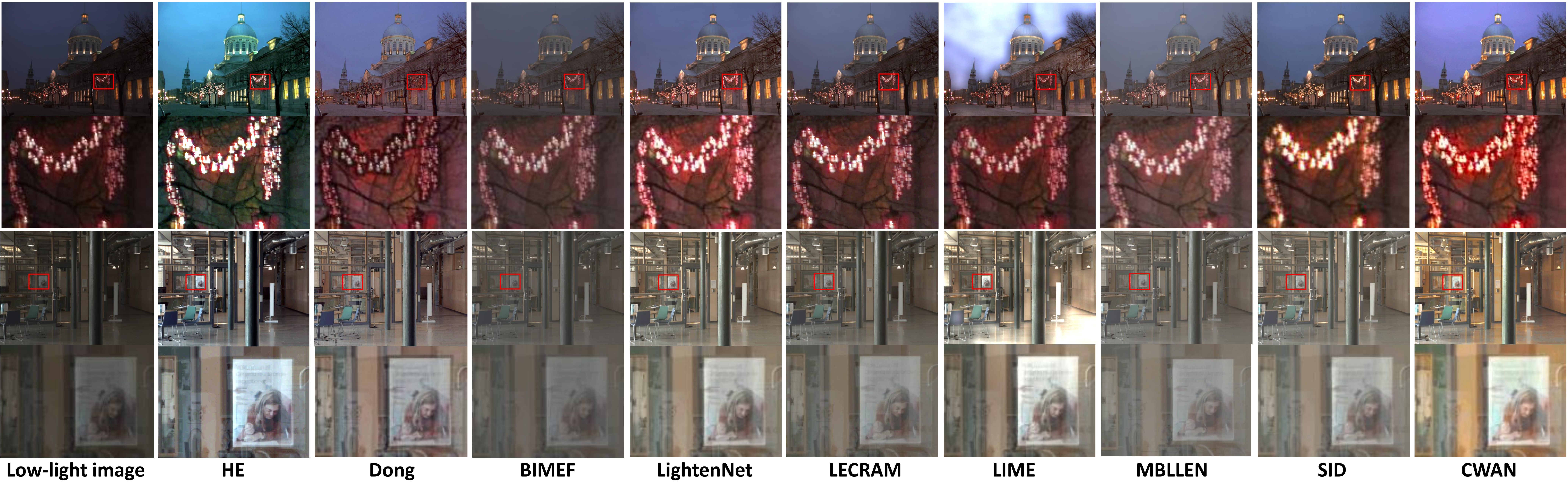}
\end{center}
\vspace{-0.4 cm}
   \caption{Qualitative results of various methods compared to CWAN on HDRDB. } 
\label{fig:Qualitative_HDR}\figvspace
\end{figure*}

\SubSection{Qualitative results}

We qualitatively compare CWAN with the same baselines in Fig.~\ref{fig:Qualitative_pascal} and~\ref{fig:Qualitative_HDR}, zooming into colorful regions where most baselines struggle in recovering the natural color. 
All of the methods tend to improve the lightness and colors of the low-light images at different performance levels.
HE usually produces over or under-enhanced regions in the image due to the increase in global contrast.
Dong, BIMEF, LECARM and Li tend to always have a grayish shade overlaid on the image.
JED and AMSR perform well on denoising, but struggle with enhancing low-light.
LightenNet enhances the brightness and contrast very well, but tends to generate white shadows very frequently, producing an unpleasant visualization.
LIME and MBLLEN method produces over-enhanced results especially in regions with original bright colors.
SID recovers from dark images remarkable well, but CWAN surpasses U-net in enhancing more vibrant colors closer to ground truth.

\Section{Conclusions}
This paper proposes a color-wise attention model for low-light image enhancement.
The method attempts to mimic the human visual system by first finding key colors in the dark image, and then spanning their attention spatially to generate a well enhanced image. 
The selected colors in the dark image are obtained by utilizing the color frequency map, and are by nature a good starting point to span the network's attention.
The experimental results reveal the advantages of our method compared to SOTA low-light enhancement methods. 
Our method can produce visually pleasing and more realistic colors similar to the ground truth images. 
\FloatBarrier

{\small
\bibliographystyle{ieee}
\bibliography{egbib}
}

\end{document}